# *Operando* Ultrafast Damage-free Diffraction-Enhanced X-ray Absorption Spectroscopy for Chemical Reactivity of Polymer in Solvents

*Zhengxing Peng[1,2]†, Antoine Lainé[1]†, Ka Chon Ng[1]†, Mutian Hua[1], Brett A. Helms[1,3], Miquel B. Salmeron[1,4]\*, Cheng Wang[1,2]\**

[1] Materials Science Division, Lawrence Berkeley National Laboratory, Berkeley, CA 94720 USA

[2] Advanced Light Source, Lawrence Berkeley National Laboratory, Berkeley, CA 94720 USA

[3] The Molecular Foundry, Lawrence Berkeley National Laboratory, Berkeley, CA 94720 USA

[4] Department of Materials Science and Engineering, University of California, Berkeley, Berkeley, CA, USA

† These authors contributed equally to this work.
\*To whom correspondence should be addressed. Email: cwang2@lbl.gov, mbsalmeron@lbl.gov

**Abstract**

Chemical recycling of plastics to its constituent monomers is a promising solution to develop a sustainable circular plastic economy. An *in-situ* X-ray absorption spectra (XAS) characterization is an important way to understand the deconstruction process. However, radiation damage, and long acquisition time, prevent such characterization for fast chemical process. Here, we present a novel experimental technique, which we name Diffraction-Enhanced X-ray Absorption Spectroscopy (DE-XAS), where the plastic material is supported on a graphene layer covering a periodic pattern of holes in a perforated $Si_3N_4$ membrane. The XAS is obtained by measuring the energy-dependent intensity of the X-ray diffracted beams going through the plastic film over the holes in the $Si_3N_4$ membrane. Our method decreases beam damage by orders of magnitude while providing good signal/noise ratio data. We demonstrate the suppression of beam damage with samples of polymethyl methacrylate (PMMA) and the *operando* characterization of polymer deconstruction process in acid with polydiketoenamine (PDK). This proof-of-concept of the DE-XAS technique shows its great potential for studying fast chemical processes, picoseconds to nanoseconds, using fast CCD detectors with short readout time. We believe the DE-XAS technique offers opportunities for studies of chemical reactions, *e.g.*, photoresist, and many others.

**Introduction**

Circular polymers with high-value monomers can be recycled from depolymerized plastic waste. This chemical recycling to monomer (CRM)[1,2] strategy has emerged as a potential solution to the development of a sustainable circular plastics economy that mitigates both the need for continuous feedstock sourcing and the accumulation of plastic waste[3]. One common CRM is depolymerization via interaction with specific solvent, such as hydrolysis or acidolysis[1], for polymers that contain carbonyl groups along the backbone[4], *i.e.*, polyesters, polyamides, and polycarbonates. One emerging and promising circular polymer, polydiketoenamine (PDK) resin, can be decomposed into its initial reusable triketone and amine monomers by hydrolysis in acid solution[5-7]. Poly(1,3-dioxolane) is also reported to be depolymerized with camphor-sulfonic acid as catalyst[8]. These depolymerization reactions can be characterized by *in situ* X-ray absorption spectra (XAS) to measure depolymerization rates[7] and its temperature dependence for an Arrhenius activation energy to understand the thermodynamics of this process.

XAS measures the photo-absorption cross section for the excitation of a core level electron to empty states of the material, which are sensitive to local chemical environment, giving a unique spectral fingerprint of the molecules. An important difficulty of X-ray studies in general is the beam damage that affects susceptible materials, like organic polymers[9]. XAS measurements require the use of a photon source with energy-tunable X-rays over a large energy range, as that currently available in synchrotron facilities. Synchrotron light sources are very bright, with a very high photon flux of $10^{12}$ photons/s/µm$^2$ and much higher. This might result in severe beam damage, manifested by bond breaking with chemical alteration of the sample. The largest contribution to damage comes from the secondary electrons generated by the excited photoelectrons in their inelastic collisions during their travel inside the material[10]. One obvious way to limit damage is to remove, or to make as thin as possible the sample and supporting substrate. A good approximation is to use ultrathin supports such as graphene, as we have recently demonstrated[11]. Another efficient way is to reduce the dose needed for the acquisition of each XAS spectra.

Here, we present a new X-ray characterization technique that both mitigates the beam damage and enables fast acquisition in the millisecond range, limited by the piezo controlled shutter speed of the monochromator in the synchrotron X-ray source. Thin sample films are deposited on an X-ray transparent support consisting of a layer graphene film covering a periodic array of holes on a 50 nm gold film-coated $Si_3N_4$ substrate. The gold film strongly attenuates the X-rays, except in the hole regions covered only by the sample and graphene. The interference of the X-rays passing through the holes produces a pattern of diffraction spots on the CCD detector, with an intensity that contains the XAS spectral fingerprint of the elements in the film. The strong diffraction intensity makes it possible to acquire the XAS with exposure time decreased by orders of magnitude while providing good signal/noise ratio data. We demonstrate the power of DE-XAS in *operando* characterizing the chemical reactivity of acidolysis of PDKs, which are an emerging family of highly recyclable polymers that can be formulated with the constituent triketone and amine monomers via "click" polycondensation reactions and undergoes hydrolysis in acid back to the triketone and amine monomers. This new technique opens the possibility for *in situ* XAS characterization for fast processing of chemical reactivities for beam sensitive materials.

**Experiment and Results**

The experimental setup used in the experiment is shown schematically in **Fig. 1a**. A cell with two 200 nm thick silicon nitride ($Si_3N_4$) membranes closing the top and bottom sides, is filled with the reactant liquid sandwiched between them. The top $Si_3N_4$ membrane perforated with a hexagonal pattern of holes is attached to a silicon frame with a 0.5 x 0.5 mm window. The holes have a diameter of 150 nm with a pitch, *i.e.*, hole-to-hole distance, of 400 nm, schematically represented in **Fig. 1b**. A thin layer of gold is evaporated on top of the perforated $Si_3N_4$ membrane, covering the $Si_3N_4$ supported regions, to attenuate X-ray from going through. In this way, the X-ray can only transmit through the hexagonal patterned holes, resulting in a strong diffraction pattern on the CCD detector due to the interference. For this purpose, the thickness of gold needs to be thick enough, *e.g.*, at least 50 nm to attenuate 80% of incoming X-rays (see **Supplementary Information**), which is crucial to generate a pattern with high contrast. A monolayer of graphene is transferred to cover the whole $Si_3N_4$ membrane (more details in **Supplementary Information**), especially the hole regions, to hold the liquid robustly. A 30 -100 nm thick film of the plastic of interest is then spin-coated with the reactant solvent added on top of it as a droplet, and then sealed with another uniform $Si_3N_4$ membrane window with no holes via epoxy applied over the edges. In transmission geometry, X-ray can only go through the hexagonal patterned holes, resulting in a hexagonal diffraction pattern. The XAS is obtained by the intensity of a certain order of diffraction peaks as a function of energy as it is scanned through an absorption edge. Diffraction peak intensities are high enough, such that millisecond exposure times are sufficient for good signal/noise ratio, which enables characterization of chemical changes in an *operando* reaction.

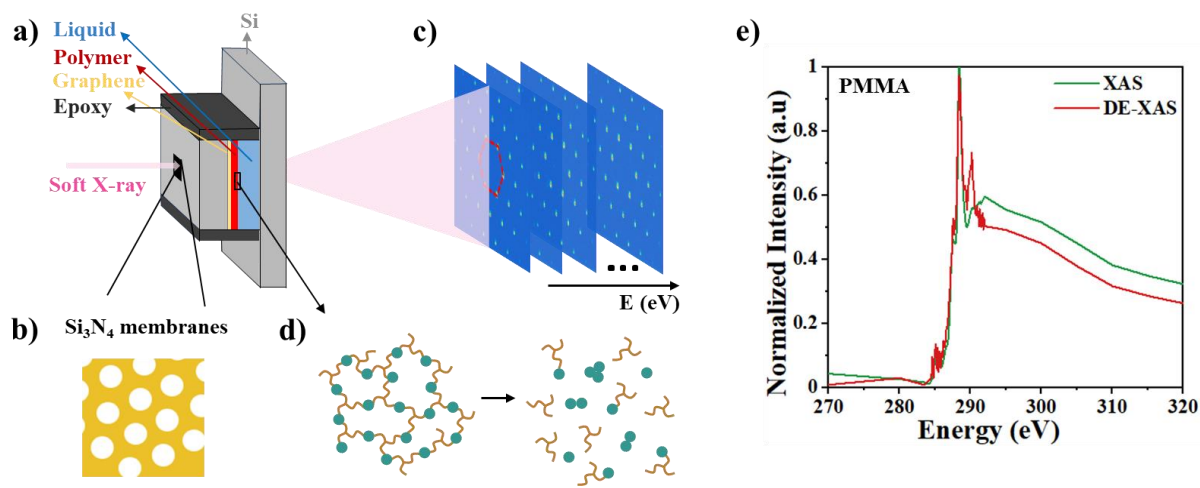

**Fig. 1 | Diffraction-Enhanced X-ray Absorption Spectroscopy (DE-XAS). a |** Schematic of the experimental setup. A liquid cell is composed of two silicon nitride ($Si_3N_4$) membrane windows sandwiching the polymer thin film on the graphene in the top gold-coated perforated $Si_3N_4$ membrane window. The liquid containing the reactant fills the space between the polymer film and the bottom $Si_3N_4$ membrane. **b |** Top $Si_3N_4$ membrane window composed of a hexagonal pattern of holes before deposition of polymer film and graphene membrane. A layer of gold of ~ 50 nm is evaporated on the perforated $Si_3N_4$ to attenuate the incoming X-rays that do not go through beyond the holes. **c |** The constructive interference of X-rays through the patterned holes produces a diffraction pattern collected with a 2D CCD. **d |** The chemical changes can be probed in an *operando* reaction. **e |** The overplot of DE-XAS of PMMA with a regular XAS spectra, where DE-XAS takes 6000 times lower dosage and 1000 times shorter exposure time.

As a test of the suppression of beam damage in DE-XAS, we use polymethyl methacrylate (PMMA) as a model system, since it is very vulnerable to beam irradiation[12,13]. First, we obtained a regular XAS of the PMMA film. The XAS is generally acquired with total electron yield (TEY) or total fluorescence yield (TFY) in reflection geometry[14] or absorption with the help of Beer-Lambert law in a transmission geometry. Here, we obtain a regular XAS of PMMA in transmission mode with the energy range of the carbon K edge, *i.e.*, 270 -320 eV. A 30nm PMMA thin film is spin coated onto a regular uniform non-perforated $Si_3N_4$ membrane window, with the transmitted X-rays collected with a photodiode. The XAS is obtained from the absorption coefficient via Beer-Lambert law, as shown in **Fig 1e,** with a total X-ray dose of

~ 300 mJ/cm² and 1 s X-ray exposure for each energy point with one XAS spectra including ~ 100 energy points. The PMMA XAS is characterized by the strong C 1s → π*$_{C=O}$ peak at ~ 288.4 eV.

For DE-XAS of PMMA, a 20 nm-thick Au layer is evaporated on the 200 nm thick Si$_3$N$_4$ perforated membrane which absorbs up to 50% of a 300 eV X-ray beam (see **Supplementary Information** for details). A 30 nm PMMA thin film was spin-coated on top of the gold-coated, hexagonally perforated substrate. In transmission geometry the incoming soft X-ray is diffracted by the hexagonal pattern of holes, forming a hexagonal pattern of diffraction spots in the CCD detector. Representative diffraction patterns, obtained by 1 ms exposure for each to energy point, are shown in **Fig. 1c**. At present, the 1 ms timescale is limited by the time required to open and close the shutter of the beamline. This limit, however, can be pushed further down using a faster shutter. The first order diffraction spots highlighted in **Fig. 1c** are used here to obtain the XAS (more details in **Supplementary Information**). The total X-ray dose for ones such DE-XAS is ~ 0.05 mJ/cm², which is 6000 times lower than the regular XAS.

The reason why diffraction can be used to provide absorption like information is that the diffraction intensity is determined by atomic scattering factor, $f = f_0 + f_1(E) + if_2(E)$, where $f_2(E)$ is proportional to the product of the X-ray absorption cross section and the X-ray energy and $f_1(E)$ and $f_2(E)$ are interrelated by Kramers-Kronig relations. It has been demonstrated by Diffraction Anomalous Fine Structure (DAFS) technique[15-17], obtaining absorption fine structure in a diffraction way. Comparing the DE-XAS with regular XAS of PMMA as shown in **Fig. 1e**, we notice some discrepancy, especially the relative intensity of the peak at ~ 290.2 eV. This discrepancy is due to the fact the DE-XAS is a result of the convolution of both $f_1(E)$ and $f_2(E)$ while the XAS is only determined by $f_2(E)$. The local chemical bond information provided by both DE-XAS and XAS agrees with each other, *e.g.*, the peak position for each kind of chemical bond.

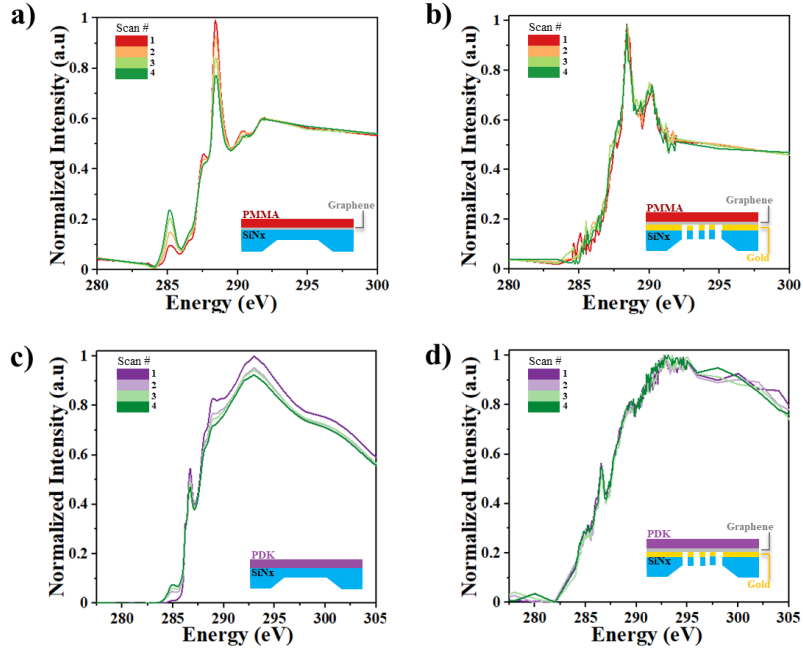

**Fig. 2 | Radiation damage observed in regular XAS measurements *vs*. the damage-free from DE-XAS measurements. a** | Regular XAS measurement for PMMA with radiation damage observed after each scan. **b** | DE-XAS measurement on PMMA with no radiation damage observed. **c** | Regular XAS measurement for PDK with radiation damage observed and **d** | DE-XAS measurement for PDK with no radiation damage observed.

To conduct an *in-situ* experiment characterizing an *operando* chemical reaction, it requires monitoring the chemical transformation by continuous XAS measurements. In PMMA case, one XAS scan takes ~300 mJ/cm$^2$, leading to radiation damage after each scan, as shown in **Fig. 2a**. A substantial chemical alteration is observed after each scan, evidenced by the decreasing intensity of the peaks located at 287.5eV, 288.4 eV and 290.2 eV, related to the C=O and C-O or C-O-C bonds[18,19]. This suggests a possible mechanism during radiation damage where the ester groups are abstracted with a reactive radical left behind, and then a carbon-carbon double bond is formed by a hydrogen abstraction[19], evidenced by the increase of the C=C peak around 285.2 eV. With DE-XAS, the radiation damage is totally suppressed with the 6000 times lower beam dosage, *i.e.*, the spectra are remain unchanged after four energy scans as shown in **Fig. 2b**. For the PDK, a circular polymer which will be studied later for the *in situ* DE-XAS characterization on the acidolysis process, similar radiation damage is observed with regular XAS measurements (**Fig. 2c**), characterized by the decrease in intensity for the peaks involving contributions from functional groups, *e.g.*, 286.6 eV for C=O bonds and 288.8 eV for exocyclic π-bonding[7], and the increasing intensity of the peak at ~ 285 eV, suggesting the formation of carbon-carbon double bonds. The mass loss is also shown by an overall decrease in peak intensity[13]. In contrast, such spectra alterations due to the radiation damage are not observed in DE-XAS experiments, as shown in **Fig. 2d**.

The *in situ* experiment starts with the deconstruction of diketoenamine (DKE) in HCl at 60 °C. DKE is a small molecule analogue of the PDK elastomers, which can be dissolved in HCl. Since its deconstruction process can be characterized by nuclear magnetic resonance (NMR), we use it as a model system to demonstrate the feasibility of DE-XAS in *operando* characterization of chemical reactivity. In HCl, the DKE can be deconstructed into triketone and amine, as illustrated in **Fig. 3a**. For NMR experiments, 5 mg of DKE was dissolved in 650 μL of 5.0 M DCl solution, transferred in a sealed-cap NMR tube, and mounted in the preheated NMR chamber. At 60 °C, the signal at 2.36 ppm (m, 1H, – NHCH(CH3)CH2–) is disappearing gradually as the deconstruction occurs as shown in **Fig. 3b** and the signal at -0.23 ppm for –CH3 is gradually shifted to -0.28 ppm as the amine is produced (**Supplementary Information Fig. S4**). The degree of conversion was calculated using the integral ratio of signal at 2.36 ppm and the signal of the –CH3 tail part of the DKE molecule, with a pseudo first-order rate law, which is determined[7] as $\sim 1.2 \times 10^{-4}$ s$^{-1}$. We also characterized the acidolysis process with regular *in situ* XAS. To prepare the liquid cell, 3 mg DKE is dissolved in 30 μL of 5.0 M HCl solution, of which a 0.3 μL droplet is sealed with two face-to-face uniform Si$_3$N$_4$ membrane windows. Then the liquid cell is mounted on a controllable heating stage for the deconstruction at 60 °C. The collected *in situ* XAS spectra is shown in **Fig. 3c**. Time-dependent density function theory (TD-DFT) can be applied to simulate the spectra for the reactants, intermediates, and products for the assignment of the characteristic XAS peaks and the spectral changes with chemical transformations. Two most important peaks are identified[7]. One prominent feature located at ~ 286.6 eV, corresponding to C=O bonds, is ascribed to either diketoenamine moieties in DKE (reactants) or to liberated ß-triketones (products). Another one is located at ~ 289.0 eV corresponding to C–N bonds which is only present in liberated amine (*i.e.*, acidolysis products), which is expected to emerge and grow as the deconstruction occurs, enabling the quantification of acidolysis rates. As shown in **Fig. 3c**, there is only a subtle growth of the peak at ~ 288.8 eV, indicating the appearance of hydrolyzed amine. In contrast, *in situ* DE-XAS, performed with the top Si$_3$N$_4$ membrane window of the above-described gold-coated and hexagonally perforated Si$_3$N$_4$ membranes, shows the much more pronounced emergence and growth of the peak at 288.8 eV, indicating deconstruction and formation of hydrolyzed amine as expected. The degree of deconstruction was monitored by observing the growing peak at 288.8 eV, as shown in **Fig. 3d**. The fitted total intensity (peak area) of the peak at 288.8 eV is assumed to follow a pseudo first-order rate law, $[Amine]_t = [Amine]_f(1 - e^{kt})$ where $[Amine]_f$ is the final concentration of amines at full deconstruction, and $[Amine]_t$ the concentration of released amines at time t, $k$ is the deconstruction rate constant, $t$ is the time elapsed for deconstruction. $k$ is calculated to be ~ $1.6 \times 10^{-4}$ s$^{-1}$, which is comparable to the NMR results. The comparison between the regular XAS and DE-XAS suggests that the diffraction enhancement also improves the spectral sensitivity to the chemical transformation, enabling the quantification of the subtle changes observed in regular XAS.

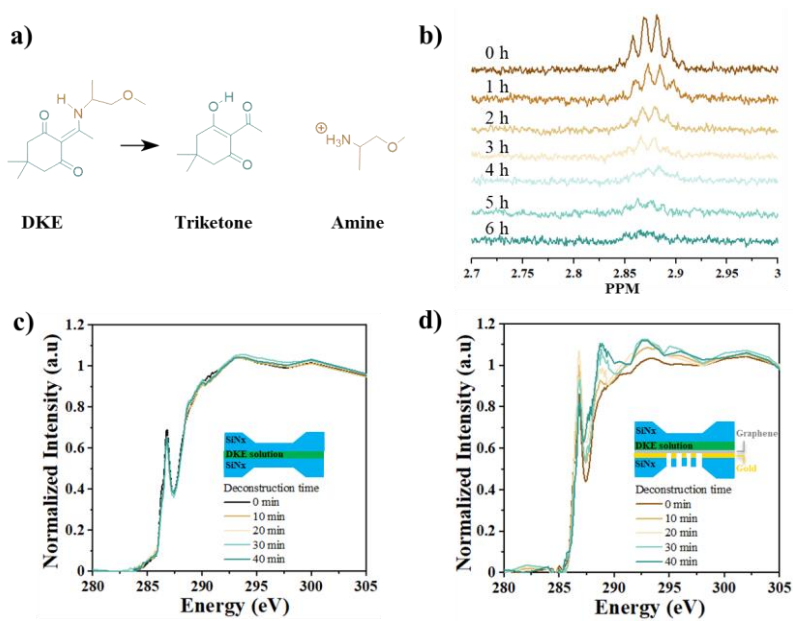

**Fig. 3 |** *In situ* **DE-XAS on the characterization of the deconstruction of DKE in HCl**. **a** | The illustration of deconstruction of DKE, a small molecule analogue of the PDK elastomers, into triketone and amine. **b** | The $^1$H NMR on the deconstruction of DKE in HCl at 60 °C. **c** | *In situ* XAS and **d** | *In situ* DE-XAS for DKE deconstruction in HCl at 60 °C.

To study the polymer-solvent interaction in a chemical reactivity with *in situ* DE-XAS, we demonstrate with the acidolysis of PDK in HCl at 60 °C after the successful characterization of the DKE deconstruction. A ~ 100 nm PDK thin film is spin coated onto the gold-coated and hexagonally perforated substrate. 0.3 μL of 5.0 M HCl solution was dropped onto the window which is covered with another uniform Si$_3$N$_4$ window and sealed with epoxy on the edge. The *in situ* DE-XAS spectra are shown in **Fig. 4b**. The deconstruction process is observed with an increase of peak at ~ 289 eV, similar to the DKE case, with the deconstruction rate calculated as ~ $6.7 \times 10^{-4}$ s$^{-1}$.

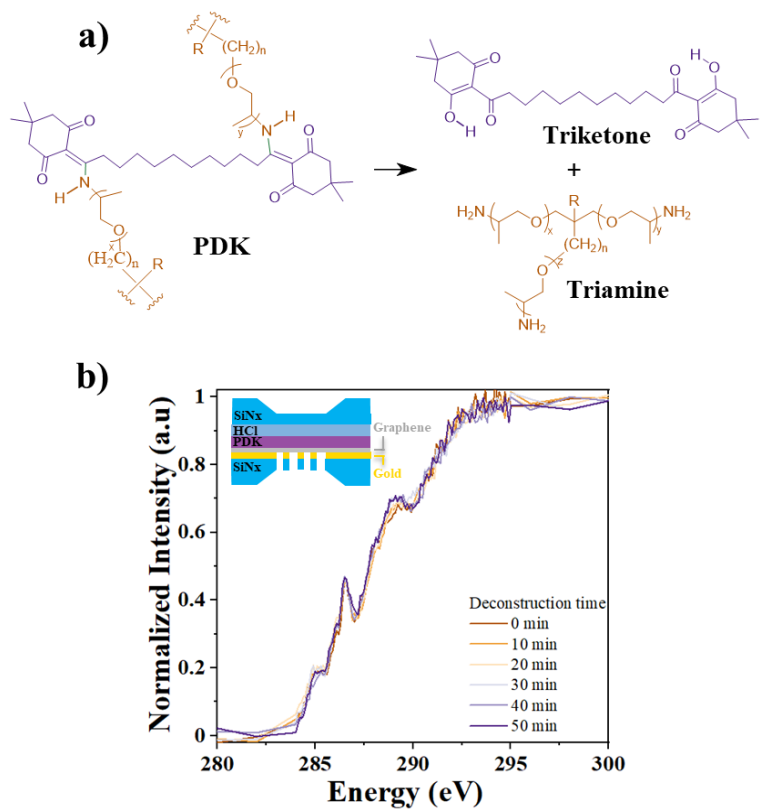

**Fig. 4 |** *In situ* **DE-XAS on the characterization of the deconstruction of PDK in HCl at 60 °C. a |** The illustration of deconstruction of PDK into triketone and amine. **b |** *In situ* DE-XAS for PDK deconstruction in HCl at 60 °C.

The depolymerization of PDK in acid studied here is a slow process occurring in minutes to hours. Its success can only demonstrate the feasibility of the *in situ* DE-XAS for *operando* characterization of the chemical transformation in a chemical reactivity. However, the DE-XAS has a much greater potential for characterizing fast processes, *e.g.*, picoseconds to nanoseconds for bond breaking reactions and the identification of the reaction intermediate[20], with a fast shutter and CCD detectors with short enough readout time. Such fast process characterization can make a breakthrough on the understanding of reaction pathways of various reactions and provide insight into the understanding of the kinetics and thermodynamics of reaction process for better polymer design or synthesis design.

**Summary**

We have developed a novel experimental technique, the Diffraction-Enhanced X-ray Absorption Spectroscopy (DE-XAS), that takes advantage of a patterned graphene substrate covering a perforated gold-coated $Si_3N_4$ membrane to achieve XAS of thin film materials with orders of magnitude decrease in beam dosage with high signal/noise ratio and short exposure time. We demonstrate the total suppression of beam damage, which is a crucial issue for XAS studies of X-ray beam susceptible polymers, with PMMA and PDK, and the feasibility of the DE-XAS in characterizing the chemical transformation in a chemical reactivity. From the scientific point of view, this technique offers the ability to accurately probe materials severely prone to beam damage. For example, photoresist materials, which are made susceptible by design to chemical alteration under irradiation, and which can be probed without additional damage from subsequent characterization. This proof-of-concept of the DE-XAS technique also show a great potential for studying fast processes, *e.g.*, picoseconds to nanoseconds for bond breaking reactions and the identification of reaction intermediates[20], with a corresponding fast shutter and CCD detectors with short enough readout time, which offers opportunities for further technical development with improved experimental setup, and paves the way for new scientific studies.

**Supporting Information**

Supporting Information is available from the publisher website or from the author.

**Acknowledgements**

This work was funded by the U.S. Department of Energy, Office of Science, Office of Basic Energy Sciences, Materials Sciences and Engineering Division under contract no. DE-AC02-05-CH11231, Unlocking Chemical Circularity in Recycling by Controlling Polymer Reactivity across Scales program CUP-LBL-Helms. Work at the Advanced Light Source was supported by the Office of Science, Office of Basic Energy Sciences, of the U.S. Department of Energy under Contract No. DE-AC02-05CH11231. Work at the Molecular Foundry and Materials Sciences Division was supported by the Office of Science, Office of Basic Energy Sciences, of the U.S. Department of Energy under the same contract. A. Laine was supported by the Inorganic/Organic Nanocomposites Program (KC3104).

# *Operando* Ultrafast Damage-free Diffraction-Enhanced X-ray Absorption Spectroscopy for Chemical Reactivity of Polymer in Solvents

*Zhengxing Peng[1,2]†, Antoine Lainé[1]†, Ka Chon Ng[1]†, Mutian Hua[1], Brett A. Helms[1,3], Miquel B. Salmeron[1,4]\*, Cheng Wang[1,2]\**

[1] Materials Science Division, Lawrence Berkeley National Laboratory, Berkeley, CA 94720 USA

[2] Advanced Light Source, Lawrence Berkeley National Laboratory, Berkeley, CA 94720 USA

[3] The Molecular Foundry, Lawrence Berkeley National Laboratory, Berkeley, CA 94720 USA

[4] Department of Materials Science and Engineering, University of California, Berkeley, Berkeley, CA, USA

† These authors contributed equally to this work.
\*To whom correspondence should be addressed. Email: cwang2@lbl.gov, mbsalmeron@lbl.gov

## Methods

**The $Si_3N_4$ membrane windows:** The patterned holey Silicon Nitride membrane widows are purchased from Ted Pella, Inc. (PELCO 21590-10. The $Si_3N_4$ membrane is 200 nm thick, the hexagonally patterned holes have a diameter of 150 nm with a pitch, *i.e.*, center-to-center distance, of 400 nm. The frame has a diameter of 3 mm with the window size of 0.5 x 0.5mm. The bottom $Si_3N_4$ membrane is made with lithography including spin-coating of photoresist, UV lithography, development, reactive ion etching (RIE) and KOH etching using a Si wafer with 100 nm low stress nitride covering both sides, purchased from Addison Semiconductor Materials, Inc. One substrate has a size of 10 x 10 mm for the Si frame and 1 x 1 mm for the $Si_3N_4$ membrane window.

**Sample preparation:** Gold is evaporated onto the $Si_3N_4$ support region for the 3 mm hexagonally perforated substrate. Graphene covered Cu foils were purchased from Graphene Factory. The backside of the Cu foil was cleaned using an $O_2$ plasma treatment. The foil was then immersed overnight in ethanol to ensure cleanliness. An Al frame was attached to untreated Cu foil side. The Cu was dissolved using a 0.2M Sodium Persulfate etching solution, and the remaining graphene sheet consolidated by the Al frame was rinsed in pure water and transferred on the gold-coated substrate and annealed under vacuum at 600K overnight. The polymer, PMMA or PDK, is spin coated on the above graphene covered substrate. PMMA is spin coated from a solution of 0.5%wt PMMA dissolved in toluene. PDK is spin-coated with dilute precursor solution prepared by dissolving 10 mg of di-topic triketone monomer and 9.4 mg of tri-topic amine monomer triamine in 1 g of THF. The spin coated sample was thermally annealed at 60 °C for 2 h to fully cure the elastomer. To make the liquid cell, 0.3 µL of solution was dropped onto the window area of the 10 × 10 mm $Si_3N_4$ window, closed over the top window, either uniform $Si_3N_4$ membrane or the hexagonally perforated substrate with gold, graphene, and polymer layer, with the window area aligned. Norland NOA68T epoxy was then applied over the edge the top smaller window. The epoxy was cured by irradiating with a handheld UV lamp for 2 min to create an air-tight seal.

**X-ray Absorption Spectroscopy (XAS).** XAS was carried out at the Advanced Light Source Beamline 11.0.1.2. Samples were mounted onto a controllable heating stage for temperature variation experiments. For the regular XAS, outcoming X-rays were collected with a photodiode (Advanced Photonix). The DE-XAS diffraction patterns were collected in vacuum on a 2D charge coupled device cooled to −45 °C (Princeton Instrument PI-MTE).

**Nuclear Magnetic Resonance Spectroscopy (NMR).** $^1$H Nuclear Magnetic Resonance Spectroscopy was carried out using a Bruker Avance II at 500 MHz. Chemical shifts are reported in δ (ppm) relative to the residual solvent peak: CDCl3: 7.26 for $^1$H. Splitting patterns are designated as s (singlet), d (doublet), t (triplet), q (quartet), and m (multiplet).

The thickness of the gold layer is critical for high contrast. Ideally, the light would come through the perforated areas and be completely absorbed everywhere else. Any light that would not be absorbed can eventually contribute to the background intensity on the detector and therefore limit the contrast of the diffraction spots. To attenuate the incoming X-ray intensity up to 80%, the gold layer needs to be at least 50 nm, as shown in **Fig. S1**.

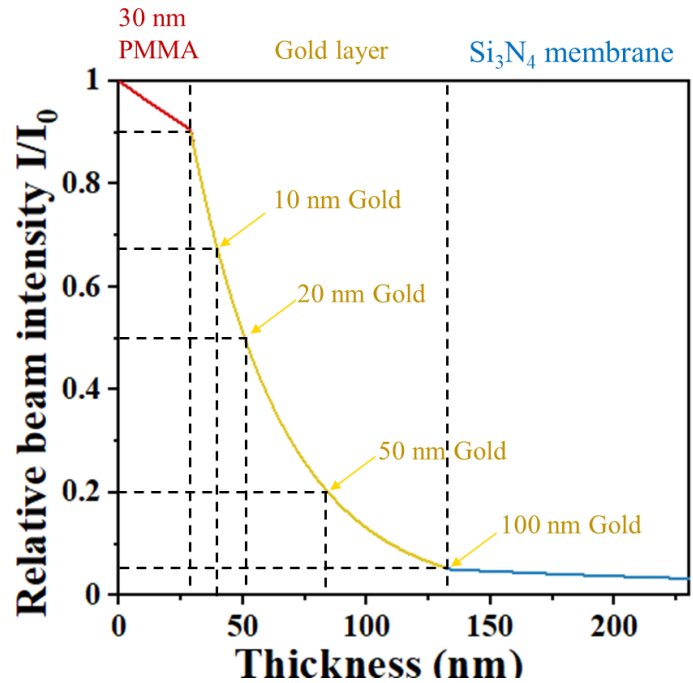

**Figure S1:** The attenuation of incoming X-ray intensity as a function of travel distance through an evaporated gold layer.

**Data processing for DE-XAS**

The diffraction patterns are acquired in a sequence of incident X-ray energies varying from 270 to 320 eV, with 1 ms exposure per data point. The XAS spectra is obtained from the added intensity of first order diffraction peaks after background subtraction, and normalized to the intensity of incident X-rays at different energies.

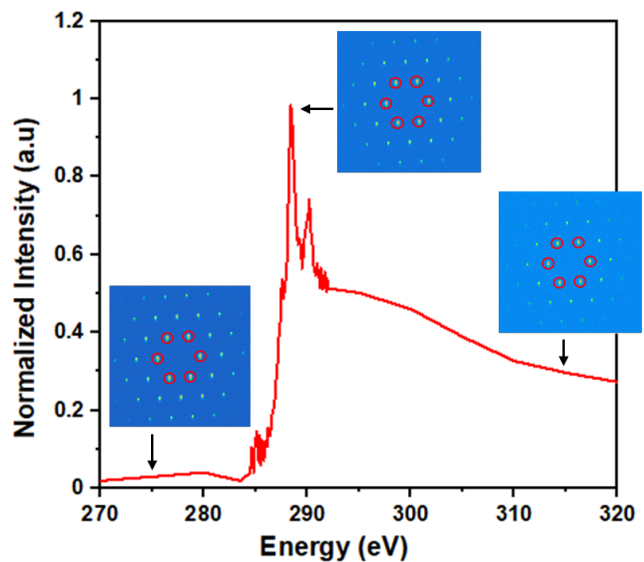

**Figure S2: From diffraction pattern to x-ray absorption spectroscopy.** Three examples of XAS spectra extracted from diffraction patterns obtained the diffraction patterns from a 30nm thick PMMA film on the hexagonal perforated substrate with incident x-ray energies of 275 eV, 290 eV, and 315 eV. The red circles highlight the first order diffraction peaks used to extract the DE-XAS spectra. Each point for XAS spectra is acquired in a sequence of incident x-ray energies varying from 270 to 320 eV, with 1 ms exposure per data point, from the intensity of the first order diffraction peaks after background subtraction.

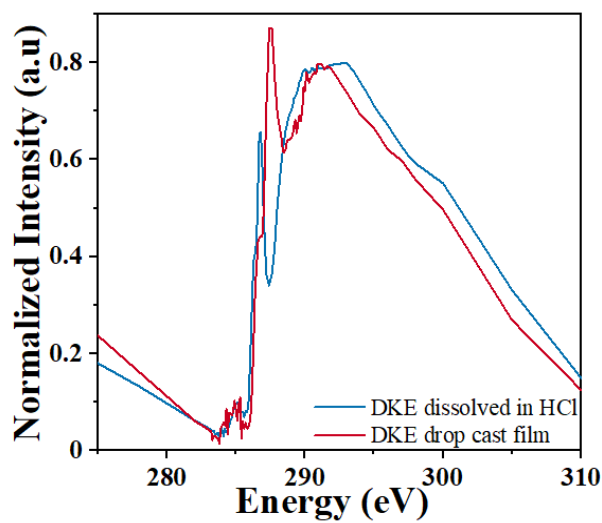

**Figure S3: XAS spectra from a DKE drop-cast film from the solution containing DKE dissolved in a HCl solution.** The deprotonated DKE contains exocyclic π-bonds that produces the strong peak at ~288.5 which does not show up when DKE is protonated.

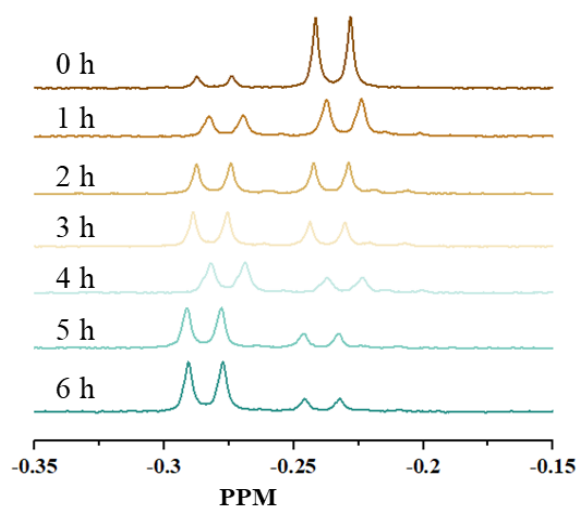

**Figure S4: NMR of DKE deconstruction in HCl at 60 °C for -CH3 group.**